\documentclass[12pt,preprintnumbers,amsmath,amssymb]{revtex4}
\usepackage{graphicx}
\usepackage{amsmath}
\usepackage{mathtools}

\begin{document}

\title{Path integral approach on Schr\"{o}dinger's cat}
\author{Zinkoo Yun}
 \email{semiro@uvic.ca}
\affiliation{Department of Physics and Astronomy University of Victoria, Canada}


\begin{abstract}
From the following thought experiments, it is demonstrated that the collapse of wave function of an isolated system is possible without external observer. 
It will be shown that the analysis by Feynman path integral method supports this conclusion. The argument is based on two assumptions: 1. The condition of Schr\"{o}dinger's cat experiment 2. Feynman path integral; This could explain Schr\"{o}dinger's cat paradox and its implication on the black hole information paradox will be discussed.
\end{abstract}


\maketitle

\section{Introduction: A clock in a box}\label{boxclock2a}

Schr\"{o}dinger's cat experiment\cite{chanik} was proposed by Erwin Schr\"{o}dinger to illustrate the absurdness of orthodox interpretation of quantum mechanics called Copenhagen interpretation which emphasis the role of the observer which collapses the wave function.
\begin{figure}
\begin{center}
    \includegraphics[height=6cm]{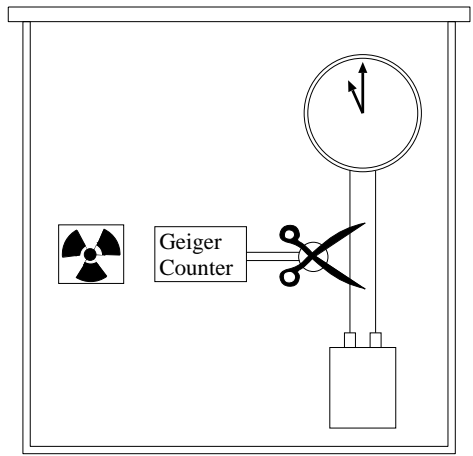}
  \end{center}
  \caption{Clock version of  Schr\"{o}dinger's cat experiment. When the Geiger counter detects decay of radio active atom, it triggers scissors to snap the power line to the clock.}
\label{clockbox_fig1}
\end{figure}

Figure \ref{clockbox_fig1} illustrates the setup of another version of Schr\"{o}dinger's cat experiment.
 We put (scissors+running clock) instead of (hammer+live cat). If the atom decays, then the Geiger counter triggers scissors to snap the power supply line to the clock.
Suppose the half life of a radio active atom is one hour. Set the time of running clock to  11:00 and cover the lid of box. Let's uncover the lid at 12:00. According to Copenhagen interpretation\cite{wigner}, before we make the observation at 12:00, the state of the clock is in quantum superposition of  $\mid\textrm{running}\rangle$ and $\mid\textrm{stopped}\rangle$ states. In many world interpretation\cite{dewitt}, the universe of coexisting $\mid\textrm{running}\rangle$ and $\mid\textrm{stopped}\rangle$ states splits into the universes of one of two states when we open the lid at 12:00. 

 Suppose we discover the clock stopped at 11:30 when we open the lid at 12:00. In this case, do we still have to believe that the collapse of wave function (or splitting world in many world interpretation) occurred at 12:00, at the moment we open the lid and make an observation as Copenhagen interpretation (or many world interpretation) insists? If the collapse of the wave function occurred at 12:00, then what should we call the physical event which occurred at 11:30?


Often it is said that if we apply quantum mechanics rigorously, we end up with the superposition of alive and dead states of a Schr\"{o}dinger's cat. In this paper, I will show that in fact we have never \emph{rigorously} applied quantum mechanics to the Schr\"{o}dinger's cat problem and will show how to do it using Feynman path integral approach.

\section{Argument about a clock in a box experiment}\label{boxclock2c}


According to Copenhagen interpretation (or many world interpretation) the wave function of an isolated system never collapse until the observer makes an observation. Thus until we open the lid, according to Copenhagen interpretation, the clock is in quantum superposition of running and stopped states and this state collapses to one of them when we observe it at 12:00.  
 Then do we always discover the clock either running or stopped at 12:00? We can easily show it does not make sense, if we introduce the second observer who observes the whole system including the first observer. 

Thus we may discover the clock stopped at 11:30 when we open the lid at 12:00.  Then we have a clear evidence that something physical event occurred at 11:30. That event must be the collapse of wave function. I cannot come up with better terminology to call that event than collapse of wave function. It means the wave function of an isolated system can collapse without external observer. In other words, the collapse of wave function of an isolated system is possible. The collapse of wave function does not require a conscious observer. 

Another possible scenario is that before we open the lid at 12:00, the state inside box is a linear combination of infinite number of quantum states corresponding to different timings of stopping clock:
\begin{equation}\label{clbx2h}
 \parbox{0.9\columnwidth}{$\mid\textrm{stopped at 11:01}\rangle+\cdots+\mid\textrm{stopped at 11:30}\rangle+ \cdots+\mid\textrm{stopped at 12:00}\rangle+\mid\textrm{running}\rangle$}
\end{equation}
So when we open the lid at 12:00, the state (\ref{clbx2h}) collapses to one of them, for example $\mid\textrm{stopped at 11:30}\rangle$;
However, in section \ref{boxclock2d}, we will prove using path integral approach that there should be no quantum superposition of $\mid\textrm{running}\rangle$ and $\mid\textrm{stopped}\rangle$ clock states \emph{in this experiment setup} (But not in general cases). So the clock has been running before 11:30 and has been stopped after 11:30, not in quantum superposition like (\ref{clbx2h}) which implies the quantum superposition of  $\mid\textrm{running}\rangle$ and $\mid\textrm{stopped}\rangle$ states before 12:00. We will discuss it detail in section \ref{boxclock2d}.

\section{Analysis by path integral}\label{boxclock2d}

\begin{figure}
\begin{center}
    \includegraphics[height=6cm]{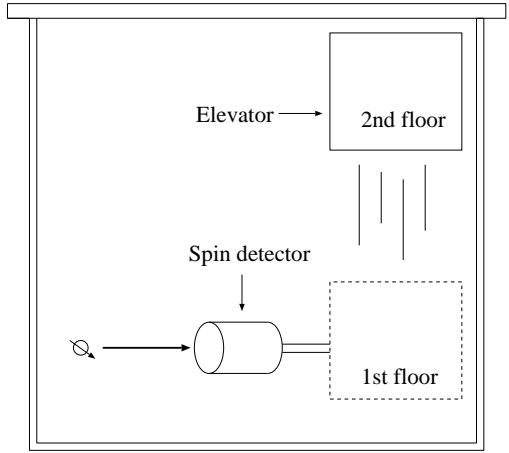}
  \end{center}
  \caption{Another version of  Schr\"{o}dinger's cat experiment. Instead of detecting decay of atom, the spin detector measures the spin $z$ direction of a spin half particle whose spin is prepared to be $\mid\uparrow\rangle_x$ state: If it measures spin up, it triggers the elevator to move from the first floor to the second floor. If it measures spin down, it triggers almost nothing.}
\label{clockbox_fig2}
\end{figure}
Instead of a Geiger counter, let's put an spin detector to measure the $z$ directional spin of a spin half particle as shown in figure \ref{clockbox_fig2}. The spin state of the spin half particle is prepared to be up in $x$ direction, $\mid\uparrow\rangle_x=1/\sqrt{2}(\mid\downarrow\rangle +\mid\uparrow\rangle)$. And instead of scissors+clock (equivalent to hammer+cat in Schr\"{o}dinger's experiment), let's put an elevator.   I introduce an elevator instead of a scissors+clock (or hammer+cat) because its motion is easier to analyze in path integral method. Then let's assume that our thought experiment works conditionally as following\footnote{The equivalent condition in Schr\"{o}dinger's cat experiment is that if the Geiger counter measures decay of atom, it triggers the hammer to break vial, if the Geiger counter does not measure decay, it does nothing.}:
\begin{equation}
\label{clbx1bx}
  \parbox{0.8\columnwidth}{If the detector measures spin up, it triggers the elevator to move from the first floor to the second floor. If the detector measures spin down, it does almost nothing.}
\end{equation}
Thus the elevator is the eventual measuring apparatus of spin state for human. If the elevator  is in the first floor, we measure the spin is down, and if the elevator is in the second floor, we measure the spin is up. In fact this is the basic process of measuring any quantum state, transforming a microscopic state to a macroscopic state which human can see. 

According to Copenhagen interpretation, before we open the lid, the state evolves smoothly from $1/\sqrt{2}(\mid\downarrow\rangle +\mid\uparrow\rangle)$  to $1/\sqrt{2}(\mid\downarrow\rangle\mid\textrm{1st}\rangle +\mid\uparrow\rangle\mid\textrm{2nd}\rangle)$.
It will be demonstrated soon that the system cannot evolve in this way according to Feynman path integral.

\begin{figure}
\begin{center}
    \includegraphics[height=7cm]{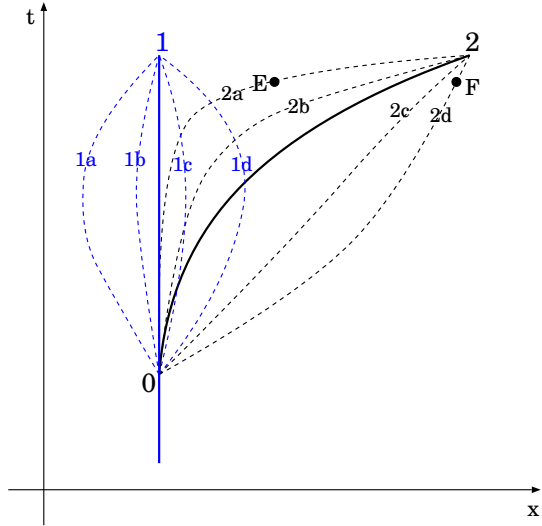}
  \end{center}
  \caption{Classical paths (solid lines) and quantum paths (dashed lines) of the experiment in figure \ref{clockbox_fig2}; Each classical path accompanies many quantum paths. The quantum paths 1a, 1b, 1c and 1d interfere themselves. The quantum paths 2a, 2b, 2c and 2d interfere themselves. However, quantum paths 1a and 2a do not interfere each other; In principle, quantum interference between two quantum paths (2a and 2d) passing two classically distinct states (Points E and F) is possible. See also the appendix \ref{boxclock2p1}.}
\label{clockbox_fig3}
\end{figure}

In path integral method, classical  Lagrangian determines all possible quantum paths.\cite{hibbs}  
\begin{equation}\label{clbx2a}
K(1,0)=\sum_{\mathclap{\textrm{all paths from 0 to 1}}} \textrm{ constant }\times e^{(i/\hbar) S[x(t)]}
\end{equation}
where $K(1,0)$ stands for the Feynman kernel from space time point 0 to 1 and $S[x(t)]$ is the action,
\begin{equation}\label{clbx2b}
S[x(t)]=\int^{t_1}_{t_0} L(\dot{x}, x, t)dt
\end{equation}
for each path $x(t)$ connecting points 0 and 1.

In figure \ref{clockbox_fig3}, the classical path 2 (2nd floor) accompanies numerous possible quantum paths (2a, 2b, 2c and 2d) and they can interfere each other. \footnote{In classical picture, the classical Lagrangian and initial position and momentum determines unique least action path. However in quantum picture, due to uncertainty of position and velocity, for given Lagrangian and initial wave function, there are many least action paths and each least action path accompanies numerous quantum paths sharing the same end points of space time which interfere themselves. Here we pick quantum paths of just two classical least action paths corresponding static and accelerating elevator for simple argument. See the appendix \ref{boxclock2p1} for more detail.} The classical path 1 (1st floor) also accompanies numerous possible quantum paths (1a, 1b, 1c and 1d) and they also interfere each other. But \emph{generally}, two quantum paths ending up two separate states do not interfere each other.
For example the quantum path 1a cannot interfere with the path 2a in figure \ref{clockbox_fig3}. Thus there is no quantum phase between the elevator in the first floor and the elevator in the second floor \emph{in the experimental setting of figure \ref{clockbox_fig2}}. 

We can understand why these two quantum paths 1a and 2a  could not interfere each other: 
If two quantum paths of an object interfere each other, the object always follows both paths simultaneously (i.e., cannot follow one or the other path) because it follows all possible quantum paths at the same time according to Feynman path integral. Suppose two paths 1a and 2a interfere each other. Then the elevator should follow both  paths simultaneously no matter what spin it measures. This contradicts the original condition (\ref{clbx1bx}) of the experiment. Thus two conditional (quantum) paths in (\ref{clbx1bx}) could not interfere each other\footnote{In Schr\"{o}dinger's cat--like experiments\cite{friedman,wineland} figure \ref{clockbox_fig4}(d), there is no direct quantum measurement on the components like (\ref{clbx1bx}) until we measure the interference effect between them.  We will discuss about it in section \ref{boxclock2e1}.};
Thus there is no phase effect between $\mid\textrm{1st}\rangle$ and $\mid\textrm{2nd}\rangle$ states and no quantum superposition is possible between them \emph{in the experimental setting of figure \ref{clockbox_fig2}}.
(In general case, the superposition between $\mid\textrm{1st}\rangle$ and $\mid\textrm{2nd}\rangle$ could  be  possible. See the section \ref{boxclock2e1}.)
The argument goes the same for the experiment in figure \ref{clockbox_fig1} and for the Schr\"{o}dinger's cat experiment.



Thus while $1/\sqrt{2}(\mid\uparrow\rangle + \mid\downarrow\rangle)$ and $1/\sqrt{2}(\mid\uparrow\rangle - \mid\downarrow\rangle)$ are different quantum states of spin, there is no distinction between $1/\sqrt{2}(\mid\textrm{1st}\rangle + \mid\textrm{2nd}\rangle)$ and $1/\sqrt{2}(\mid\textrm{1st}\rangle - \mid\textrm{2nd}\rangle)$ or between  $1/\sqrt{2}(\mid\textrm{running}\rangle + \mid\textrm{stopped}\rangle)$ and $1/\sqrt{2}(\mid\textrm{running}\rangle - \mid\textrm{stopped}\rangle)$ or between  $1/\sqrt{2}(\mid\textrm{dead}\rangle + \mid\textrm{alive}\rangle)$ and $1/\sqrt{2}(\mid\textrm{dead}\rangle - \mid\textrm{alive}\rangle)$ \emph{in our experimental settings}. This explains why the superposition of $\mid\textrm{dead}\rangle$ \emph{and} $\mid\textrm{alive}\rangle$ cat is not just improbable but impossible \emph{in Schr\"{o}dinger's cat experiment}.

Thus, as oppose to many other quantum interpretations, we are not allowed to express the state right after interaction as such
\begin{equation}\label{clbx2c}
\frac{1}{\sqrt{2}}\Big(\mid\downarrow\rangle\mid\textrm{1st}\rangle + \mid\uparrow\rangle\mid\textrm{2nd}\rangle\Big)
\end{equation}
\emph{in the experiment of figure \ref{clockbox_fig2}} or
\begin{equation}\label{clbx2c1}
\frac{1}{\sqrt{2}}\Big(\mid\textrm{decay}\rangle\mid\textrm{stopped}\rangle + \mid\textrm{not decay}\rangle\mid\textrm{running}\rangle\Big)
\end{equation}
\emph{in the experiment of figure \ref{clockbox_fig1} }  or
\begin{equation}\label{clbx2d}
\frac{1}{\sqrt{2}}\Big(\mid\textrm{decay}\rangle\mid\textrm{dead}\rangle + \mid\textrm{not decay}\rangle\mid\textrm{alive}\rangle\Big)
\end{equation}
\emph{in Schr\"{o}dinger's cat experiment} because the expression (\ref{clbx2c}), (\ref{clbx2c1}) and (\ref{clbx2d}) imply the quantum superposition between two resulting states.
Thus if we discovered the clock stopped at 11:30 when we open the lid at 12:00 in the experiment described in figure \ref{clockbox_fig1}, we can say that the clock has been running before 11:30 and has been stopped after 11:30, not in superposition of $\mid\textrm{running}\rangle$ and $\mid\textrm{stopped}\rangle$ like (\ref{clbx2h}).
This supports the conclusion of previous section -- The collapse of wave function of an isolated system is possible without external observer.

The point is that if $\mid\textrm{dead}\rangle$ and $\mid\textrm{alive}\rangle$  are on two quantum paths interfering each other, then the cat always should follow both  paths at the same time no matter detecting decay of atom or not (according to Feynman path integral). This contradicts the original condition of Schr\"{o}dinger's cat experiment.  It means two conditional paths \emph{in Schr\"{o}dinger's cat experiment} (ending up $\mid\textrm{dead}\rangle$ / $\mid\textrm{alive}\rangle$), should not interfere each other. So there is no phase effect between $\mid\textrm{dead}\rangle$ and $\mid\textrm{alive}\rangle$, thus no quantum superposition between $\mid\textrm{dead}\rangle$ and $\mid\textrm{alive}\rangle$ \emph{in the experiment setup proposed by Schr\"{o}dinger}. (But not in general case. See the section \ref{boxclock2e1}.)

\section{Schr\"{o}dinger's cat--like experiments}\label{boxclock2e1}

\begin{figure}
\begin{center}
    \includegraphics[height=8cm]{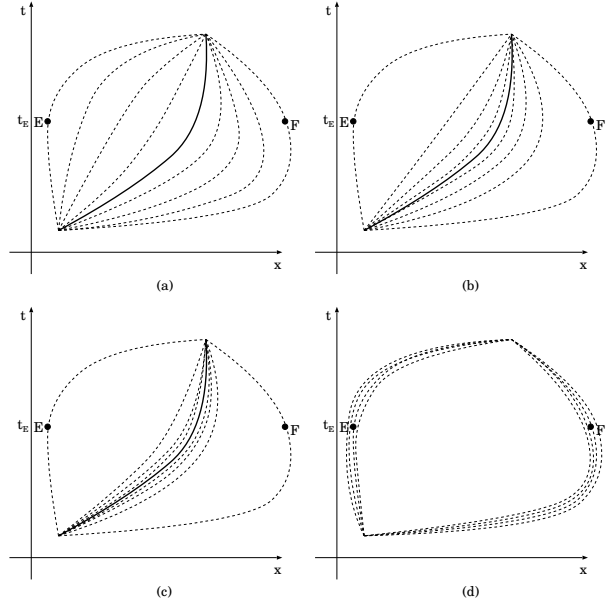}
  \end{center}
  \caption{Classical paths (solid lines) and quantum paths (dashed lines). (a) When the classical action is comparable size to the Plank constant $h$. There is not single dominating quantum path, so its interference effect is clear to be detected. (c) When the classical action is much bigger than $h$. Quantum paths adjacent to the least action path (classical path) are the only dominating bunch.  So the interference effect of other paths (For instance passing E and F) is not clear to be detected and ignorable. See also the appendix \ref{boxclock2p1}. (d) In Schr\"{o}dinger cat--like experiments\cite{friedman,wineland}, two components of a wave function split into two quantum paths passing macroscopically distinct states (Points E and F). There are two dominating bunch of quantum paths. So their interference effect is detectable.}
\label{clockbox_fig4}
\end{figure}

We cannot directly observe two eigenstates in quantum  superposition at once even for a microscopic object (For example, We cannot directly observe an electron in two position at the same time because we cannot measure two eigenvalues at the same time.). We can measure only the effect of quantum superposition. We confirm quantum superposition by measuring interference effect (phase effect) between quantum paths.\footnote{Quantum interference is between quantum paths rather than between static quantum states.} 

Though we cannot directly observe it, a microscopic object, such as a photon, easily can be in a quantum superposition of two classically distinct states (points E and F in figure \ref{clockbox_fig4}(a)) for a given classical Lagrangian of the object, in the sense we can observe its quantum interference effect easily; However as the size (the classical action compare to Plank constant) of the object gets bigger, it gets more and more improbable for it to be in quantum superposition between two classically distinct states, in the sense it gets harder to observe its quantum interference effects (Figure \ref{clockbox_fig4}(b)(c))\cite{hibbs}. 

As shown in figure \ref{clockbox_fig4}(c) (and appendix \ref{boxclock2p1}), a macroscopic object, such as an elevator, is most likely in superposition of quantum states which are almost indistinguishable classically. Still, in principle, a macroscopic object may evolve to superposition of many classically distinguishable quantum states (Points E and F in figure \ref{clockbox_fig4}(c))  whose quantum effect is almost ignorable most of time; 
Therefore we understand that there is no such a boundary between quantum and classical world. Any object evolves from one quantum state to another quantum state guided by Lagrangian of it.
\\

I don't insist that \emph{in general}, two quantum paths passing $\mid\textrm{dead}\rangle$ and $\mid\textrm{alive}\rangle$ (or $\mid\textrm{running}\rangle$ and $\mid\textrm{stopped}\rangle$) could not interfere each other: We can imagine splitting of components starting at the same space time point, passing points 1 and 2 and ending up at another same space time point in figure \ref{clockbox_fig3}. If there is no direct measurement of these components until the end point, then these two quantum paths can interfere each other. So the quantum superposition between $\mid\textrm{dead}\rangle$ and $\mid\textrm{alive}\rangle$ is possible\footnote{In appendix \ref{boxclock2p1}, it is shown that we can achieve the superposition of first and second floor states of an elevator just by waiting.  Simple calculation tells us that we never achieve that superposition in our lifetime.}; \footnote{However, arguing whether the superposition of $\mid\textrm{dead}\rangle$ and $\mid\textrm{alive}\rangle$ is possible or not maybe pointless as long as we can show it is in principle not achieved by the experiment proposed by Schr\"{o}dinger.}

There were some experiments to realize the superposition of two macroscopic distinct states using laser cooled trapped ions\cite{wineland} and using the superconducting quantum interference device (SQUID)\cite{friedman}\footnote{Friedman. {\it et al.} could detect the difference between $1/\sqrt{2}(\mid\textrm{clockwise current}\rangle + \mid\textrm{anti clockwise current}\rangle)$ and $1/\sqrt{2}(\mid\textrm{clockwise current}\rangle - \mid\textrm{anti clockwise current}\rangle)$.} (figure \ref{clockbox_fig4}(d)). 

In both Schr\"{o}dinger's cat--like experiments, there is no direct measurement on each components (Points E and F in figure \ref{clockbox_fig4}(d)) until we measure the interference effect between them. Thus the object could follow simultaneously both paths and they interfere each other. In other words there is a quantum phase effect between two component states; On the other hand, in Schr\"{o}dinger's cat experiment, there is a direct quantum measurement on $\mid\textrm{decay}\rangle$ and $\mid\textrm{not decay}\rangle$ before measuring interference effect between them.  So there is no phase effect between $\mid\textrm{decay}\rangle$ and $\mid\textrm{not decay}\rangle$, hence no phase effect between  $\mid\textrm{dead}\rangle$ and $\mid\textrm{alive}\rangle$ states.

\section{Loose of information or deterministic universe?}\label{boxclock2e}

It is widely accepted that any isolated quantum system, no matter how complicated, evolves smoothly by unitary evolution process. This picture is consistent all experimental results though it leads to the measurement problem. Besides, this picture implies that the state of clock in figure \ref{clockbox_fig1} is in superposition of running and stopped states before opening the lid which is disproved in section \ref{boxclock2d} by Feynman path integral.

On the other hand, the picture in this paper says that the isolated system evolves by the combination of smooth unitary process and collapsing of wave function. (i.e., The clock may have been stopped long before opening the lid.) which seems more free to the measurement problem. In fact, both pictures predict the same experimental results.
\\

We saw that the collapse of wave function of an isolated system is possible.
In classical picture, the Brownian motion is random motion with probability of Gaussian distribution. So the motion of molecule is indeterministic. However if we include environment, the evolution of isolated system is believed to be deterministic; In quantum picture, there seems to be indeterministic collapses of wave function within an isolated system without observer. Would this collapse to be also a kind of unitary evolution if we count the change of environment?
If the collapse of wave function is a kind of unitary process, then it implies that the universe is deterministic which tastes is not that sweet. On the other hand, if the collapse is true random process, the wave function collapse is fundamental. Then it implies that the information  may be  destroyed by pure quantum process.

\section{Summary}\label{boxclock2f}
The thought experiment proposed in figure \ref{clockbox_fig1} (or with a spin detector) suggest that the collapse of wave function of an isolated system may occur long before any external observer makes an observation. Thus it demonstrates that the collapse of wave function of an isolated system is possible. 
With path integral interpretation, this could resolve the Schr\"{o}dinger's cat paradox with no boundaries between quantum and classical world: 

According to Feynman path integral, if $\mid\textrm{dead}\rangle$ and $\mid\textrm{alive}\rangle$  are on two quantum paths interfering each other, then the cat should always follow both paths at the same time (i.e., cannot follow one or the other path). So the cat must follow both  path simultaneously no matter detecting decay of atom or not. But this contradicts the original condition of Schr\"{o}dinger's cat experiment.  It means two conditional paths \emph{in Schr\"{o}dinger's cat experiment} (ending up $\mid\textrm{dead}\rangle$ / $\mid\textrm{alive}\rangle$), do not interfere each other. So there is no phase effect between $\mid\textrm{dead}\rangle$ and $\mid\textrm{alive}\rangle$, thus no quantum superposition between $\mid\textrm{dead}\rangle$ and $\mid\textrm{alive}\rangle$ \emph{by the experiment setup proposed by Schr\"{o}dinger}.
(In general case, the quantum superposition between $\mid\textrm{dead}\rangle$ and $\mid\textrm{alive}\rangle$ could be possible. We analyzed why Schr\"{o}dinger cat--like experiments\cite{friedman,wineland} could demonstrate the superposition of two macroscopically distinct states while Schr\"{o}dinger's cat experiment itself could not.). 

If the collapse of wave function is fundamental, then it implies that the information  may be  destroyed by pure quantum process. This may give a clue to resolve the black hole information paradox.

\section{Further study}\label{boxclock2g}
We have seen that the wave function of an isolated system evolves by combination of unitary processes and collapses of wave function. What is the detail mechanism of it? Is the collapse of wave function fundamental? In what condition does the collapse take place? Is it something to do with irreversibility of quantum states of measuring apparatus? Is the timing of collapse itself random by some probability? Further study is needed.

\section*{Acknowledgments}
I would like to thank Dr. Werner Israel for his advises and useful comments on this work.

\appendix
\section{Multiple least action paths due to uncertainty}\label{boxclock2p1}
\begin{figure}
\begin{center}
    \includegraphics[height=7cm]{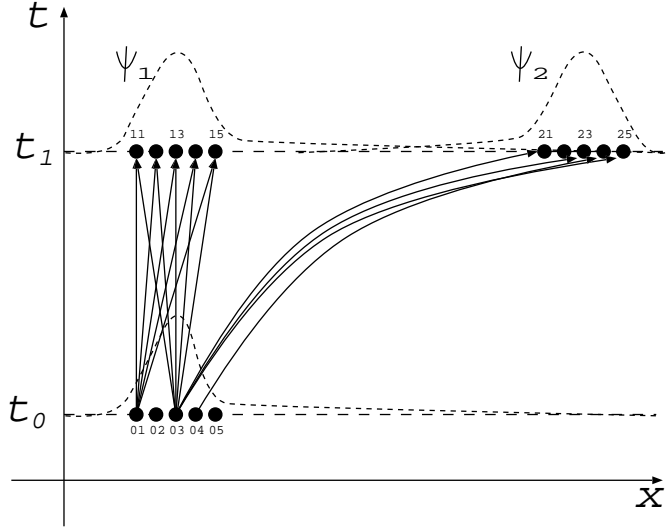}
  \end{center}
  \caption{Multiple least action paths of macroscopic apparatus due to quantum uncertainty. Even static object has multiple least action paths.}
\label{clockbox_fig5}
\end{figure}
In path integral, we cannot draw unique classical path from the information about the Lagrangian and the initial point of space time, because it does not have single momentum value due to quantum uncertainty: In figure \ref{clockbox_fig5}, I draw the motion of an elevator in space time. The initial stationary elevator evolves to $\psi_1$ if there is no change of the classical Lagrangian of the elevator. Due to uncertainty, the initial position of the elevator is represented by the Gaussian distribution. The points 01, 02, 03, 04 and 05 represent few selected positions of the elevator. The arrow connecting 03 and 13 does not represent unique least action path even we know the form of Lagrangian (in this case Lagrangian of free particle), because of uncertainty of momentum. So depending on the velocity of the elevator, arrows from 03 to any points (11, 12, 13, 14 and 15) can be the classical paths. And each of these classical path accompanies numerous quantum paths sharing  the same start and end points which interfere themselves; The argument goes the same for the arrows from 01 to any points (11, 12, 13, 14 and 15). Besides, we have uncertainty of position represented by the Gaussian distributions.

However, if the object is macroscopic like an elevator, the uncertainty of position and velocity is usually quite small. So the classical path connecting (03,11) and the classical path connecting (03,15) are very close to each other. The classical path connecting (01,11) and the classical path connecting (01,15) are also very close to each other. And the uncertainty of the initial position also very small compare to its size for the macroscopic object like an elevator. So the points 01, 02, 03, 04 and 05 are very close to each other compare to the size of the elevator.Thus we can effectively say that almost all classical paths of static elevator are very close to the path connecting 03 and 13 (i.e., the least action path in classical picture):
Suppose the uncertainty of position and velocity of an object  at $t_0$ is $\Delta x_0$ and $\Delta v_0$. Then from the uncertainty relation of Gaussian wave packet $m\Delta v_0\Delta x_0=\hbar$, we find the quantum uncertainty $\Delta x_{v}$ at $t_1$ due to $\Delta v_0$:
\begin{equation}\label{clbx2i}
\Delta x_v=\Delta v_0 \Delta t=\frac{\hbar\Delta t}{m\Delta x_0}
\end{equation}
where $\Delta t=t_1-t_0$.
Combining with the uncertainty $\Delta x_0$, the uncertainty $\Delta x_1$ at $t_1$ is
\begin{equation}\label{clbx2j}
\Delta x_1=\sqrt{\Delta x^2_0+\Delta x^2_v}=\Delta x_0\sqrt{1+\Big(\frac{\hbar\Delta t}{m\Delta x^2_0}\Big)^2}
\end{equation}
For an electron with $\Delta x_0$ equal to the Compton wavelength $h/m_e c\sim 10^{-11}\,\mathrm{m}$ of electron,
$\hbar/m\Delta x^2_0\sim 10^{19}/\mathrm{s}$, while for an elevator with the same $\Delta x_0$ and $m=1\,\mathrm{kg}$, $\hbar/m\Delta x^2_0\sim 10^{-12}/\mathrm{s}$. Thus while the position uncertainty of an electron grows from the size of Compton wave length of an electron to 100,000 km in a second, that of an elevator hardly grows. 

In short, because of quantum uncertainty, any object actually follows multiple least action paths and numerous quantum paths accompanying each of lease action path, and in case of macroscopic object, all of these least action paths are very close to the classical least action path which we can find uniquely in classical picture; The argument goes the same for an accelerating elevator (to $\psi_2$). There are many least action paths for given an initial point and given Lagrangian (with potential term). However because the uncertainty of position and velocity are very small for an elevator we can effectively say that almost all these least action paths of an accelerating elevator are very close to the classical least action path connecting 03 and 23.

For a static elevator, in principle, we may find the elevator at the point 23 of $\psi_1$ at $t_1$ according to equation (\ref{clbx2j}). With this respect we may say that the elevator is in superposition of 1st and 2nd floors. But this probability is quite small for macroscopic measuring device like the clock in figures \ref{clockbox_fig1} or the elevator in figure \ref{clockbox_fig2} within our life time scale. We can ignore this tail effect so that we can trust macroscopic measuring device; In order to make the static elevator in the superposition of 1st and 2nd floor practically, we just leave the elevator alone for quite a while without disturbing, so that its uncertainty spread significantly over 1st and 2nd floor. Simple calculation from (\ref{clbx2j}) shows that with the initial uncertainty $\Delta x_0$ equal to the Compton wavelength of electron, we have to wait at least $\sim 10^{23}$ seconds in order to observe the quantum superposition of first and second floor states of an elevator. This also explains why we don't observe a quantum tunneling effect in everyday life, even if we assume a system decoupled from its environment. 

On the other hand, if the measuring device is microscopic then the overlapping between $\psi_1$ and $\psi_2$ could be quite significant within short time according to equation (\ref{clbx2j}).  Thus when we find the microscopic device at the point 23, we cannot be sure whether this is due to change Lagrangian or just due to uncertainty. In other words, it is hard to confirm the wave function collapse of an object with microscopic measuring apparatuses. Then we cannot trust microscopic measuring devices.

\end{document}